\documentclass[a4paper]{jpconf}
\usepackage[numbers,sort&compress]{natbib}
\usepackage{graphicx}
\usepackage[caption=false]{subfig}     
\usepackage{siunitx}
\usepackage{amsmath}
\usepackage{amssymb}
\graphicspath{ {./figures/} }
\usepackage{hyperref}
\hypersetup{ 
	colorlinks=true,
	linkcolor=blue,
	citecolor=blue,
	urlcolor=blue}

\begin{document}
\title{Subgrid moving contact line model for direct numerical simulations of bubble dynamics in pool boiling of pure fluids}

\author{Vadim S. Nikolayev$^1$, Linkai Wei$^2$ and Guillaume Bois$^2$}

\address{$^1$ Universit\'e Paris-Saclay, CEA, SPEC, CNRS, 91191 Gif-sur-Yvette Cedex, France}
\address{$^2$ Universit\'e Paris-Saclay, CEA, Service de Thermo-hydraulique et de Mécanique des Fluides, 91191 Gif-sur-Yvette Cedex, France}
\ead{vadim.nikolayev@cea.fr}

\begin{abstract}
This contact line vicinity model is conceived as a subgrid model for the DNS of bubble growth in boiling. The model is based on the hydrodynamic multiscale theory and is suitable for the partial wetting case. On the smallest length scale (distance from the contact line) $\sim\SI{100}{nm}$, the interface slope is controlled by the Voinov angle. It is the static apparent contact angle (ACA) that forms due to evaporation, similarly to previous models neglecting the contact line motion. The calculation of the Voinov angle is performed with the generalized lubrication approximation and includes several nanoscale effects like those of Kelvin and Marangoni, vapor recoil, hydrodynamic slip length and interfacial kinetic resistance. It provides the finite values of the heat flux, pressure and temperature at the contact line. The dynamic ACA is obtained with the Cox-Voinov formula. The microscopic length of the Cox-Voinov formula (Voinov length) is controlled mainly by the hydrodynamic slip. The integral heat flux passing through the contact line vicinity is almost independent of the nanoscale phenomena, with the exception of the interfacial kinetic resistance and is mostly defined by the dynamic ACA. Both the dynamic ACA and the integral heat flux are the main output parameters of the subgrid model, while the local superheating and the microscopic contact angle are the main input parameters. The model is suitable for the grid sizes $>\SI{1}{\mu m}$.
\end{abstract}

\section{Introduction}\label{sec:intro}

In nucleate boiling, the vapor bubble nucleation and growth on the heating wall favors high heat fluxes from the surface to the liquid thus providing a high rate of overall heat exchange. The bubble growth modelling is a longstanding challenge and is still debated in the literature. In the modeling of bubble growth, the line of triple vapor-liquid-solid contact (CL) presents a separate issue because of its geometrical singularity that has two main consequences for its vicinity loosely called microregion. First, high rates of the heat and mass exchange in the microregion need to be included into the model to reproduce correctly the wall cooling at the CL  \cite{Sodtke06} and bubble growth rate \cite{Torres24}. Second, strong hydrodynamic flows caused by the mass exchange cause strong variation of interface slope at the bubble base (called apparent contact angle $\theta_{app}$) \cite{Park23}. It needs to be correctly modeled as it presents the geometrical boundary condition that affects the bubble shape.

Several versions of microregion models have been developed since the pioneering works \cite{Wayner,Stephan}, most of them considering the static CL and the complete wetting case where the long-range molecular interaction are implemented through the disjoining pressure. However, in most practical cases of boiling, the wetting is partial and CL moves. The microregion model for the partial wetting case has been developed by Jane\v{c}ek and Nikolayev \cite{EuLet12,PRE13}. It includes several nanoscale effects like those of Kelvin and Marangoni, vapor recoil, hydrodynamic slip length and interfacial kinetic resistance. Since such a model provides the finiteness of temperature, pressure and heat flux, its numerical implementation is robust. Its version suitable for description of large $\theta_{app}$ has been developed based on the extended lubrication approach \cite{Mathieu_thes,IPHT14,JFM22}. The analysis with the moving CL \cite{PRE13movingCL,SWEP22,JFM22} shows that, while the interface shape is defined by evaporation-induced liquid flow at the distance $\lesssim\SI{100}{nm}$ from CL, the flow caused by the CL motion is dominant at larger length scales. Thanks to such a scale separation, the effects of evaporation and CL motion can be considered separately. Elaboration of the subgrid model for bubble growth simulations based on this idea is the objective of this study.


\begin{figure}[t]
	\centering
	\includegraphics[width=7cm]{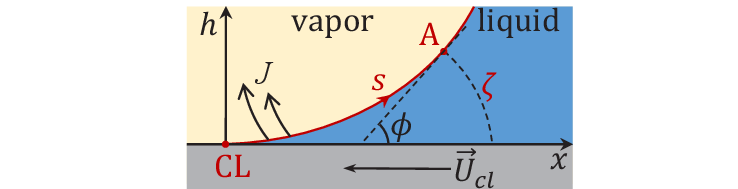}
	\caption{Schematics of the curved interface in the contact line (CL) vicinity in the reference of the immobile CL (the wall is moving with the speed $U_{cl}$).}
	\label{fig:wedge}	
\end{figure}

\section{Generalized lubrication approach}\label{sec:Lubrication}

Since the microregion size is typically much smaller than the CL radius, the subgrid problem can be solved in 2D, even for DNS of the 3D or axisymmetric bubble. Instead of operating with the Cartesian coordinates  ($x,h$) of a point A belonging to the interface (Fig. \ref{fig:wedge}), one uses a curvilinear coordinate $s$ that runs along the interface, with $s=0$ at the CL. Therefore, the following geometrical relations hold:
\begin{equation}\label{param}
\begin{split}
{\mathrm{d} h}/{\mathrm{d} s}&=\sin \phi,\\
{\mathrm{d} x}/{\mathrm{d} s}&=\cos \phi,
\end{split}
\end{equation}
where $\phi$ is the local interface slope. For the point A, one introduces a straight wedge formed by the tangent to the interface and the wall and the corresponding circular arc passing though A with the length
\begin{equation}\label{zeta}
\zeta=h\phi/\sin\phi.
\end{equation}
The generalized lubrication equation, written in the reference where CL is immobile but the wall moves with the CL velocity $U_{cl}$ in the opposite direction, reads \cite{JFM22}
\begin{equation}\label{eq:GEA}
\frac{\mathrm{d} }{{\mathrm{d} s}}\left\{ {\frac{1}{{\mu G(\phi )}}\left[ \frac{\zeta}{2}( \zeta + 2 l_s )\frac{\mathrm{d} \sigma }{\mathrm{d} s}+\frac{\zeta^2}{3}(\zeta+3l_s)\frac{\mathrm{d}\Delta p}{\mathrm{d} s} \right] - {U_{cl}}\zeta \frac{{F(\phi )}}{{G(\phi )}}} \right\} =  - \frac{J}{\rho_l},
\end{equation}
where $\Delta p$ is the interfacial vapor-liquid pressure jump, $J$, the evaporation mass flux, $l_s$, the hydrodynamic slip, $\sigma$, the surface tension, $\mu$, liquid shear viscosity, $\rho_l$, the liquid density, and
the coefficients
\begin{align}
F(\phi)&=\frac{2\phi^2}{3}\frac{\sin{\phi}}{\phi-\sin\phi\cos\phi},\label{eq:F}\\
G(\phi)&=\frac{\phi^3}{3}\frac{4}{\sin\phi\cos\phi-\phi\cos2\phi}\label{eq:G}
\end{align}
are the corrections to the conventional lubrication theory satisfying $F(\phi\to 0)=G(\phi\to 0)=1$. The surface tension gradient is \begin{equation}\label{eq:MarEff}
  \frac{\mathrm{d} \sigma }{\mathrm{d} s} \simeq  - \gamma \frac{\mathrm{d} T^i}{\mathrm{d} s},
\end{equation}
where $T^i$ is the interfacial temperature and $\gamma=-\mathrm{d}\sigma/\mathrm{d} T$ is generally positive for pure fluids. The interface shape is defined with the equation
\begin{equation}\label{eq:pJumpPr(s)}
\frac{\sigma}{\cos\phi}\frac{\mathrm{d}^2h}{\mathrm{d} s^2} -\Delta p = {J^2}\left( \frac{1}{\rho_v} - \frac{1}{\rho_l} \right),
\end{equation}
that involves the vapor recoil; $\rho_v$ is the vapor density. The heat flux is determined as
\begin{equation}\label{eq:J(s)}
  q = {k(T_w-T^i)}/{\zeta},
\end{equation}
where $k$ is the liquid thermal conductivity. The mass flux is $J=q/{\cal L}$, where ${\cal L}$ is the latent heat. The wall temperature $T_w=T_{sat}+\Delta T$ differs from the saturation temperature $T_{sat}$ for the given system pressure by the superheating $\Delta T$. The vapor recoil also affects the interfacial temperature
\begin{equation}\label{eq:TintK-pr-Ri}
T^i = T_{sat}\left[ 1 + \frac{\Delta p}{{\cal L}\rho_l}  + \frac{J^2}{2{\cal L}}\left( \frac{1}{\rho_v^2} - \frac{1}{\rho_l^2} \right) \right] +  R^i q.
\end{equation}
The interfacial thermal resistance (unity accommodation coefficient is assumed) reads
\begin{equation}
\label{eq:Ri}
    R^i = \frac{T_{sat} \sqrt{2\pi R_v T_{sat}} (\rho_l-\rho_v) }{2 {\cal L}^2 \rho_l \rho_v },
\end{equation}
where $R_v$ is the specific vapor constant.
The boundary conditions for these equations are defined by the geometry at the contact line. At the CL ($s=0$), the geometry implies
\begin{equation}\label{eq:BC}
\begin{split}
& h= 0, \\
&\phi= \theta_{micro},
\end{split}
\end{equation}
where $\theta_{micro}$ is the static contact angle measured under adiabatic conditions. More precisely, it should be static receding angle for $U_{cl}>0$, and static advancing angle in the opposite case. Another boundary condition of zero interface curvature
\begin{equation}\label{Kcl}
\frac{{\partial  \phi }}{\partial s}= 0
\end{equation}
is given at the point M that is the right end of the integration interval $s=s_M$, which is, typically, commensurate with the DNS mesh size, 5--\SI{20}{\mu m}.
The set of lubrication equations (\ref{eq:GEA}, \ref{eq:pJumpPr(s)}) is of fourth order, so one more boundary condition is required. The temperature continuity along the interfaces at the CL $T^i=T_w$ is necessary so $J(s\to 0)$ is bounded. From \eqref{eq:TintK-pr-Ri}, one obtains the condition
\begin{equation}\label{p00}
\Delta p=\frac{{\cal L}\rho_l}{T_{sat}}(\Delta T-R^iJ{\cal L})-\frac{J^2\rho_l}{2}\left(\frac{1}{\rho_v^2}-\frac{1}{\rho_l^2}\right),
\end{equation}
valid  as $s\to 0$. The asymptotic analysis \cite{JFM22} of the above model yields
\begin{equation}\label{eq:Jcl}
J(s\to 0)=\frac{U_{cl}F(\theta_{micro})}{\dfrac{G(\theta_{micro})}{\theta_{micro}\rho_l}+\dfrac{l_s{\cal L}\theta_{micro}}{\mu k}\gamma}.
\end{equation}
The pressure value at the CL is easily obtained by substituting \eqref{eq:Jcl} into \eqref{p00}. One can see that the pressure is finite at the CL.

Within such an approach, the power per unit CL length supplied to the microregion is
\begin{equation}\label{Qmicro}
Q_{micro}=\int_0^{s_M}q\mathrm{d}s.
\end{equation}
This value is one of four output parameters of the subgrid model. The second parameter is $\theta_{app}=\phi(s=s_M)$. The other two parameters are $x_M$ and $h_M$ (typically, of the order of the grid size), i.e. the Cartesian coordinates of the point M obtained as the solutions of Eqs. \eqref{param} for $s=s_M$. As for the input parameters, in addition to the fluid parameters, $l_s$ and $\theta_{micro}$ (which are typically all constant for a given simulation), they are two: the superheating $\Delta T$ and the CL velocity $U_{cl}$. The strategy of the integration of the subgrid model into the DNS is discussed in \cite{Wei24}.
\begin{figure}[hb]
\centering
\subfloat[$\theta_{app}$ as a function of superheating and CL velocity] {\includegraphics[width=7cm,clip]{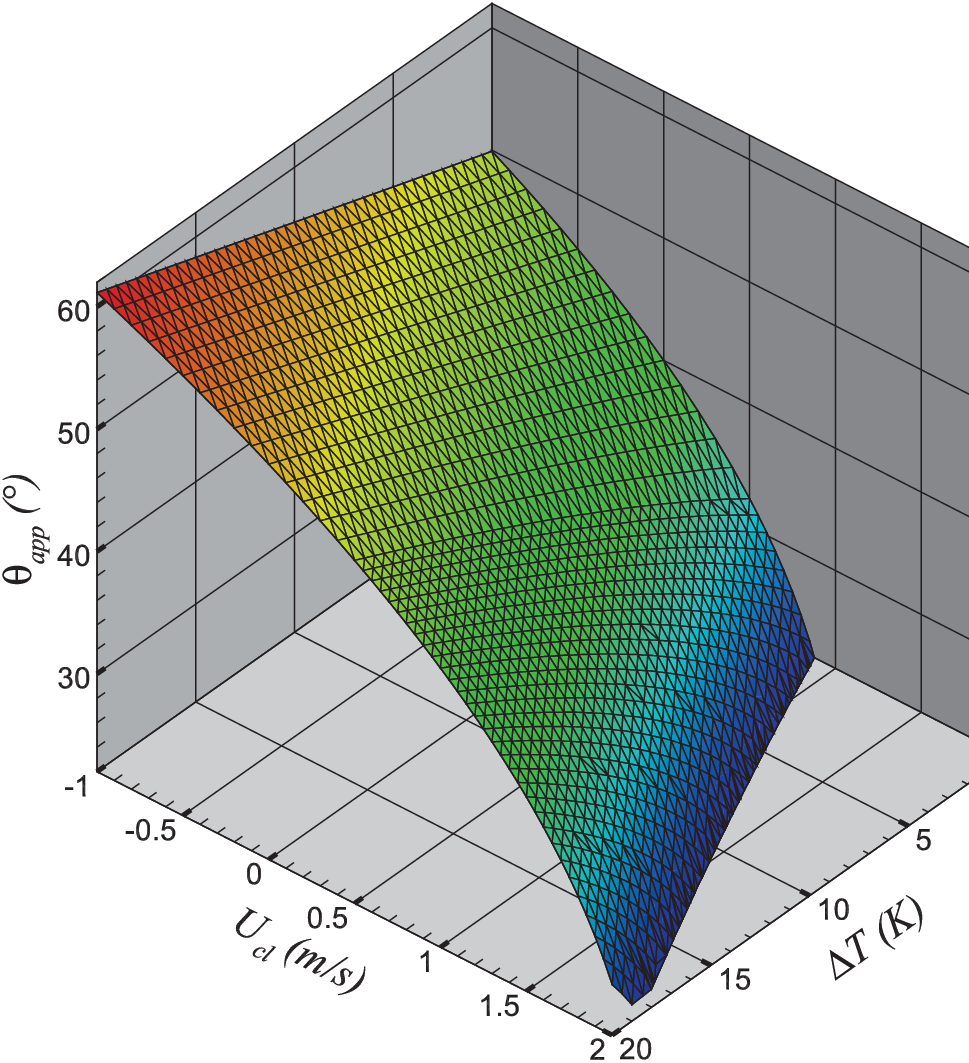}%
\label{thetaApp_vCLDT_40}}\hspace*{5mm}%
\subfloat[$\theta_{app}$ and $Q_{micro}$ at $U_{cl}=\SI{0.5}{m/s}$ as functions of $\Delta T$. The solid lines show the results of the problem of \autoref{sec:Lubrication}, while the dotted lines correspond to the approach of \autoref{sec:multi}. The $\theta_V$ variation is also given as a dashed line.
] {\includegraphics[width=8cm]{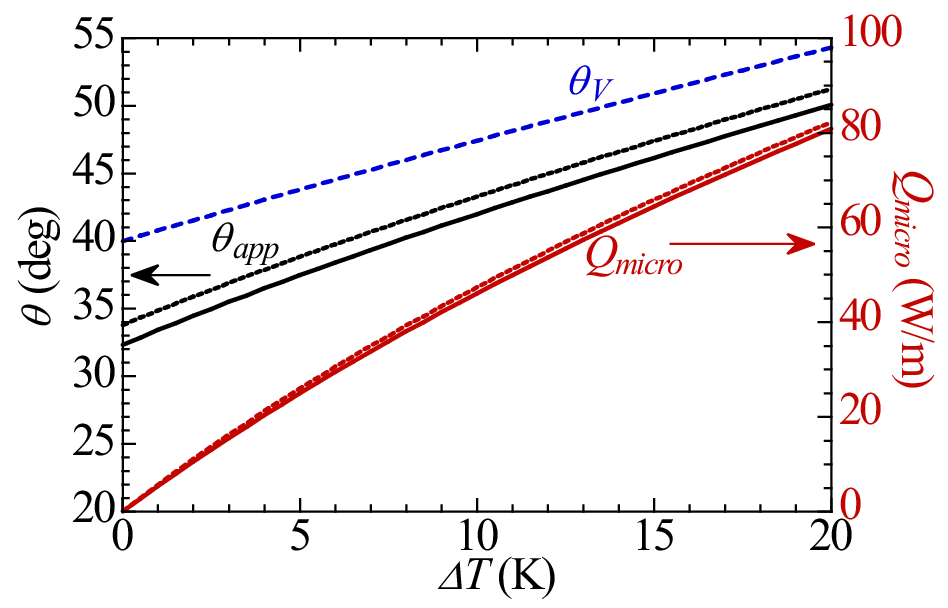}%
\label{ThetaAppQmicro}}
\caption{The subgrid problem results computed for water at atmospheric pressure; $s_M=\SI{10}{\mu m}$, $l_s=\SI{10}{nm}$, and $\theta_{micro}=40^\circ$.}\label{fig:thetaApp}
\end{figure}

The problem posed above is discretized by using the finite volume discretization scheme. The grid is chosen to be progressively denser when approaching CL. The system of equations is solved as linear, the nonlinear terms are treated with iterations. All the fluid parameters mentioned above are assumed to be constant. An example of $\theta_{app}$ variation resulting from such a problem is given in Fig.~\ref{thetaApp_vCLDT_40}.

\section{Multiscale semi-analytical approach}\label{sec:multi}

\begin{figure}[hb]
\centering
\includegraphics[width=7cm,clip]{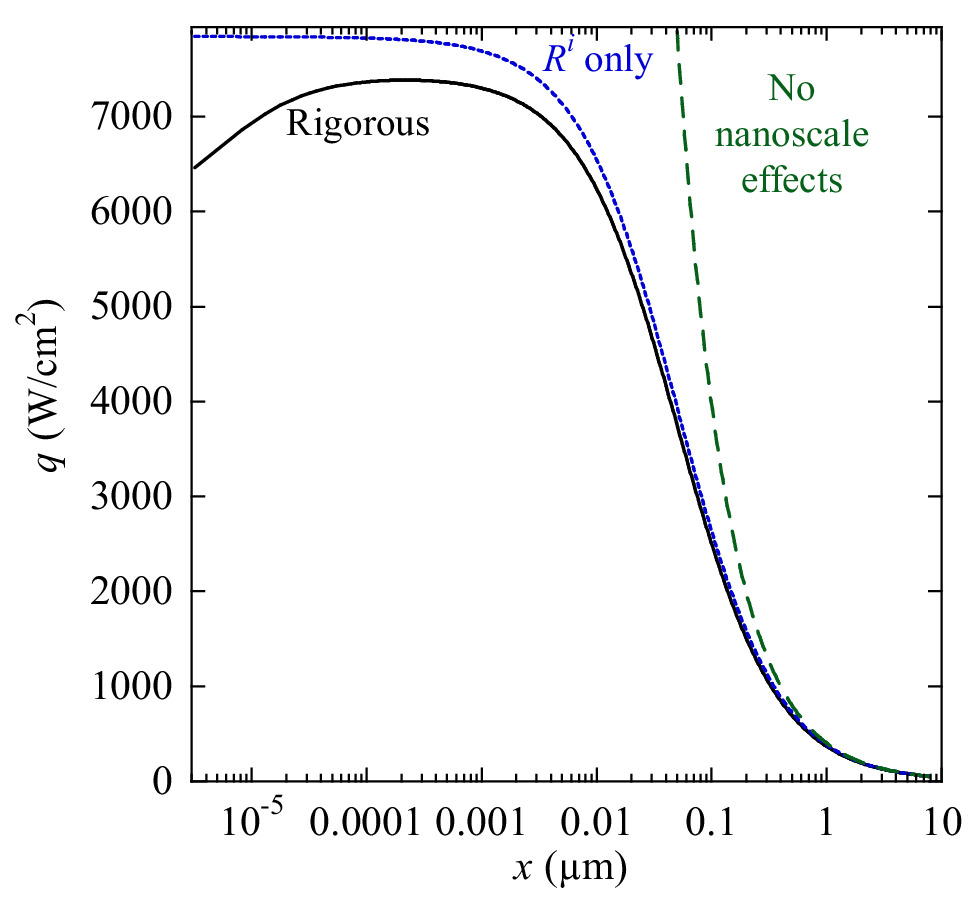}
\caption{$q(x)$ variation calculated for the same parameters as Fig.~\ref{ThetaAppQmicro} and for $\Delta T=\SI{5}{K}$.}\label{fig:q}
\end{figure}

The above calculation is quite delicate numerically because of several reasons. First, it is the necessity to treat a large range of the length scales at the same time (Fig.~\ref{fig:q}). Second, the convergence of the above algorithm is quite slow  at small $\theta_{micro}$. A simpler method can be used to obtain the same output parameters. As mentioned in the introduction, there is a scale separation between the evaporation-induced liquid flow and the flow caused by the CL motion. While the evaporation defines the interface curvature in a region of a length $\ell_V\sim\SI{100}{nm}$ (called the Voinov length) from the CL, the CL motion controls the interface shape at a larger (but still micrometric) distance. Note that the interface slope provided by the evaporation only is called the Voinov angle $\theta_V$. It can be obtained within a theory presented in \autoref{sec:Lubrication} corresponding to $U_{cl}=0$ (dashed line in Fig.~\ref{ThetaAppQmicro}). This quantity is similar to the ACA obtained with all the previous microregion models \cite{Stephan,Torres24}. The ACA accounting for the CL motion can be obtained \cite{PRE13movingCL,SWEP22,JFM22} with the generalized Cox-Voinov equation
\begin{equation}\label{eq:Cox-Voinov}
  \theta_{app}^3 = \theta_V^3-9Ca\ln \frac{x_M}{\ell_V},
\end{equation}
where $Ca=\mu U_{cl}/\sigma$ is the capillary number. This equation corresponds to the solution of the CL motion problem without evaporation satisfying the boundary condition \eqref{Kcl}. The Voinov length was shown to be controlled by the slip length:
\begin{equation}\label{eq:lVslip}
\ell_{V}\simeq\frac{3l_s}{e\theta_V},
\end{equation}
where $e=\exp(1)$. To obtain  $\theta_{app}$, the multiscale approach consists in solving Eq. \eqref{eq:GEA} without the last term in the l.h.s. thus getting $\theta_V$, and then use Eq.~\eqref{eq:Cox-Voinov}. The deviation of Eq.~\eqref{eq:Cox-Voinov} from the rigorous solution is as small as $\sim 1.5^\circ$ (Fig.~\ref{ThetaAppQmicro}). One mentions a stronger deviation of $\theta_{app}$ from $\theta_V$ (dashed line) that shows the necessity to account for the CL motion.

To propose a simplified approach for the calculation of $Q_{micro}$, one approximates the interface as a straight wedge with the opening angle $\theta_{app}$ so
\begin{equation}\label{xhAnal}
\begin{split}
&x\simeq s\cos \theta_{app},\\
&h\simeq s\sin \theta_{app},
\end{split}
\end{equation}
These approximations provide a high enough accuracy and, if used for $s=s_M$, define the output parameters $x_M,h_M$ of the subgrid model. One can now compare the heat flux variation resulting from \autoref{sec:Lubrication} and a ``$R^i$ only'' formula
coming from Eqs.~(\ref{eq:J(s)}, \ref{eq:TintK-pr-Ri}) where all the other nanoscale effects are neglected:
\begin{equation}\label{qliNo}
q = \frac{k\Delta T}{s\theta_{app}+R^ik}.
\end{equation}
The corresponding ``$R^i$ only'' curve is displayed in Fig.~\ref{fig:q}. The difference with the rigorous calculation of the \autoref{sec:Lubrication} model with Eq.~\eqref{qliNo} exists only for tiny $s$. They cause a negligible contribution to the integral \eqref{Qmicro}, which explains an extremely weak difference between the numerical approach (solid red line in Fig.~\ref{ThetaAppQmicro}) and the formula (dotted red line)
\begin{equation}\label{QmicroNo}
Q_{micro}= \frac{\Delta T\lambda_l}{\theta_{app}}\ln\left(\frac{s_M\theta_{app}}{R^i\lambda_l}+1\right)
\end{equation}
obtained by integration of the flux \eqref{qliNo}. This feature has been previously mentioned in \cite{Huber17}. It should be mentioned however that for a smaller $\theta_{micro}$, the microregion contribution becomes more important because the fluxes are larger and the influence of microscopic effects extends over larger distances. For the sake of comparison, we display in Fig.~\ref{fig:q} also the $q(x)$ dependence without the account of nanoscale effects (Eq. \eqref{qliNo} with $R^i=0$, dashed curve) that results in the infinite $Q_{micro}$.

\section{Conclusion}
We propose a subgrid model for inclusion into the DNS describing the moving contact line problem with evaporation. Two approaches are compared: a rigorous calculation based on the generalized lubrication approximation and a semi-analytical model. The latter is much more economical as it requires a computational study with only one parameter, the local superheating at the contact line. The other parameter, the contact line velocity, is accounted for by using an analytical expression based on the multiscale argument. The semi-analytical approach provides a sufficient accuracy for the evaluation of the output parameters of the subgrid model, which are the apparent contact angle, heat power per unit contact line length and corresponding microregion sizes.

\bibliography{ContactTransf}

\end{document}